\documentclass[conference]{IEEEtran}
\makeatletter
\def\ps@headings{%
\def\@oddhead{\mbox{}\scriptsize\rightmark \hfil \thepage}%
\def\@evenhead{\scriptsize\thepage \hfil \leftmark\mbox{}}%
\def\@oddfoot{}%
\def\@evenfoot{}}
\makeatother
\usepackage{cite}
\usepackage{amsmath,amssymb,amsfonts}
\usepackage{algorithmic}
\usepackage{graphicx}
\usepackage{textcomp}
\usepackage{xcolor}
\usepackage{physics}
\usepackage{balance}

\usepackage{scalerel}
\def\pitslab{\kern0em\raise-.01em\hbox{\scalerel*{\includegraphics{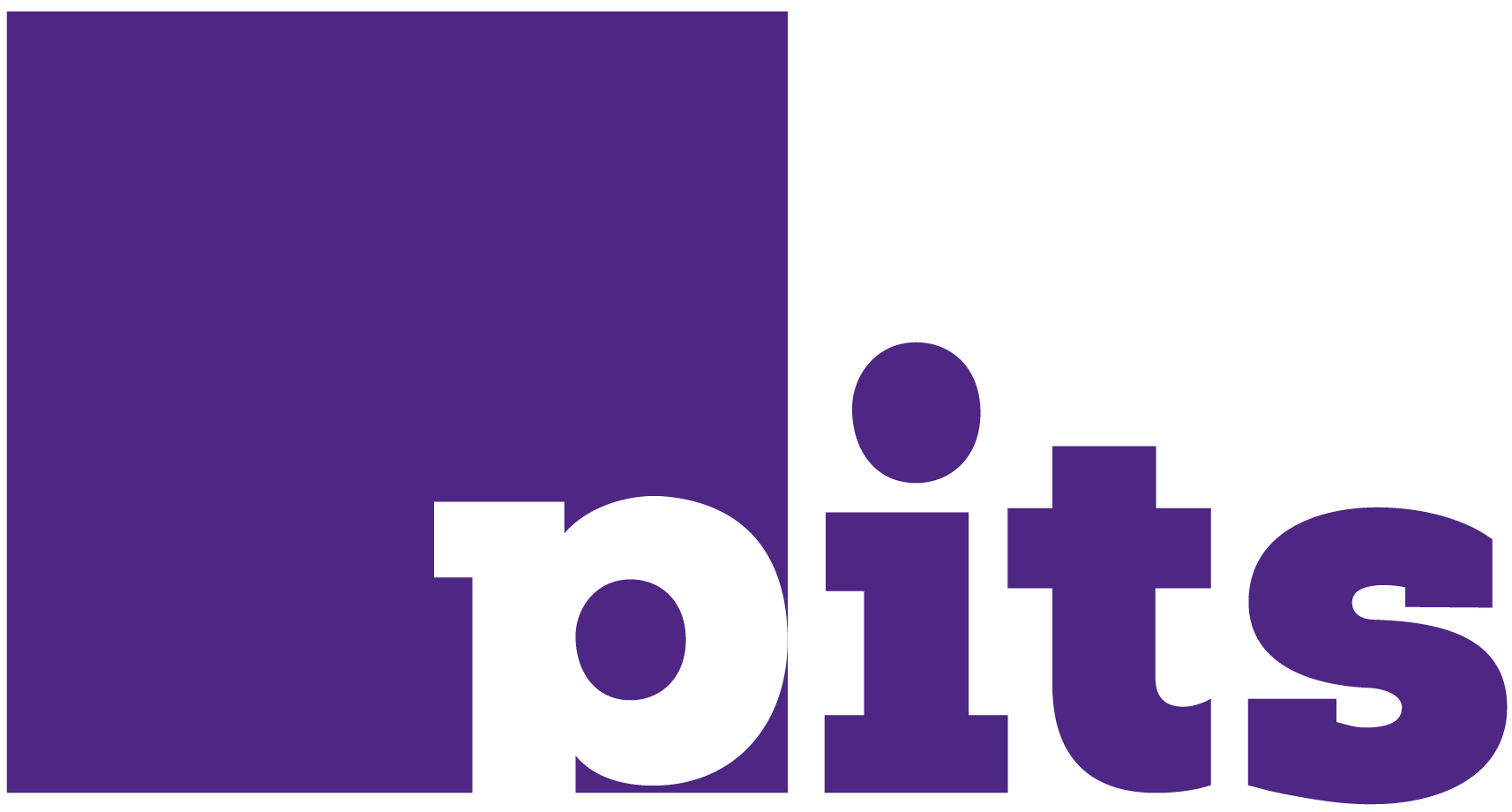}}{X}}}

\newcommand{\FISHER}{\mathcal{I}}
\newcommand{\EXPECT}{\mathrm{E}}
\newcommand\numberthis{\addtocounter{equation}{1}\tag{\theequation}}
\def\SIGMA(#1,#2){\Sigma_{#1 \cup #2}}

\def\BibTeX{{\rm B\kern-.05em{\sc i\kern-.025em b}\kern-.08em
    T\kern-.1667em\lower.7ex\hbox{E}\kern-.125emX}}
    
\begin{document}

\title{A Practical Approach to Navigating the Tradeoff Between Privacy and Precise Utility\\
}

\author{\IEEEauthorblockN{Chandra Sharma\qquad George Amariucai}
\IEEEauthorblockA{\textit{Department of Computer Science} \\
\textit{Kansas State University}\\
Manhattan, Kansas \\
\{ch1ndra, amariucai\}@ksu.edu}
}

\maketitle

\begin{abstract}

Due to the recent popularity of online social networks, coupled with people's propensity to disclose personal information in an effort to achieve certain gratifications, the problem of navigating the tradeoff between privacy and utility attracted a lot of recent interest and generated a rich body of research. A critical prerequisite to solving the problem is to appropriately capture the privacy and the utility aspects in the problem formulation. Most of the existing works' focus is on the notion of privacy, while utility loss is often treated as the undesirable but necessary distortion of the true data, introduced by the privacy mechanism. By contrast, we are interested in modelling utility differently, by associating it with specific attributes of a user, just like privacy is associated with specific private attributes in the literature. Our model of utility facilitates a better and more precise privacy mechanism, and achieves better privacy-utility tradeoffs. We further incorporate into our problem formulation a practical constraint on acceptable loss in utility per unit gain in privacy, which allows users to customize the privacy mechanisms in order to account for the relative values that each user associates with their own privacy and utility. This paper discusses the intricacies of our utility model and the corresponding privacy-utility tradeoff, and introduces a heuristic greedy algorithm to solve the problem.
\end{abstract}


\pagestyle{empty}
\section{Introduction}
The effective use of online services, such as search engines, social networks, e-commerce and streaming services, often requires that we share our data with these services. The shared data is typically used by the service providers to enhance the utility of the services. For instance, e-commerce services use the information about a user's purchases to recommend new products that may be of interest to the user. Similarly, streaming services use a user's ratings of various movies to recommend new and potentially interesting movies to the user. Consequently, massive amount of individual data is used by the service providers which can be subject to malicious use, often via seemingly innocuous release to other parties.

The biggest concern with the shared data is the privacy of the data. The shared data may contain various private attributes about a user such as sex, race, sexual orientation, political affiliation etc. Ideally, it is desirable that no private information is leaked from the shared data while, at the same time, the data is as useful as it could be. Unfortunately, the privacy requirements and the utility requirements are often contradictory. In the literature, this problem is most commonly formulated as a privacy-utility tradeoff problem and various researchers have come up with various application-specific solutions (see, for instance, \cite{sankar2013utility, makhdoumi2013privacy, rajagopalan2011smart, alvim2011differential}).

An important part of the problem formulation is to sufficiently capture the intuitive understanding of privacy and utility. A common information disclosure pattern typically involves a randomization mechanism which takes some input data and produces a perturbed output. The goal is to hide sensitive information from the data while maintaining the perceived utility. In essence, any privacy metric should nontrivially relate between the disclosed data and the sensitive information. In the literature, such relation is often captured using privacy metrics such as \textit{Differential Privacy} \cite{alvim2011differential, yin2017location, yang2012differential}, \textit{Correlated Differential Privacy} \cite{zhu2014correlated, wu2017game}, \textit{Mutual Information} \cite{makhdoumi2013privacy, rajagopalan2011smart, sankar2013utility}, \textit{Changes in min-entropy} \cite{braun2009quantitative, smith2009foundations, alvim2011differential}, \textit{Fisher Information} \cite{farokhi2017fisher, farokhi2017optimal, wang2018preserving}, \textit{Maximal Leakage} \cite{issa2016operational, liao2017hypothesis} and \textit{Maximal Correlation} \cite{li2018maximal, asoodeh2015maximal, makhdoumi2013privacy}. The choice of a particular privacy metric depends on the use case; if it is desirable to hide the identity of users in a database, metrics such as \textit{Differential Privacy} are adequate whereas if the goal is to hide specific private attributes of a user, information-theoretic metrics such as \textit{Mutual Information} or \textit{Maximal Leakage} are more useful.

As with the case of privacy, it is also critical to mathematically capture the intuitive understanding of utility. Loosely speaking, utility can be thought of as a meaningful use of the shared information. The subject of the utility could either be the person who shares the information or the party that uses the shared information, or both. In any case, the better the information is used, the higher the utility which makes utility highly reliant on the quality of the information. Any randomization mechanism, therefore, essentially decreases the utility of the information. Utility is commonly quantified using a distortion function (see, for instance, \cite{sankar2013utility}, \cite{makhdoumi2013privacy}, \cite{rajagopalan2011smart} and \cite{erdogdu2015privacy}) which, roughly speaking, captures an overall loss of information between the data before and after randomization. Such a formulation of utility makes an important distinction from privacy: privacy is considered an \textit{individual} concept whereas utility is considered an \textit{aggregate} concept \cite{li2009tradeoff}. Unfortunately, such formulation of utility is often unnecessary, and even insufficient, when utility is considered on individual basis and can be associated with specific attributes of a user just like privacy. In such cases, it is undesirable to capture utility with an aggregate function, instead, utility should relate to each individual's utility attributes which allows for much better privacy-utility tradeoff. Further, the existing works seem to overlook the fact that in many cases, it may be unacceptable to sacrifice a significant amount of utility just for a negligible gain in privacy, even if the randomization ensures that the end utility is higher than a preset value. Of course, a na\"ive fix is to set a higher value for the desired utility but that would, at other times, negatively impact the end privacy gain. Instead, a potential fix is to let the users who share their data in return of some utility decide the acceptable gain in privacy per unit loss in utility, on individual basis. To the best of our knowledge, such a formulation has not been discussed in the existing literature. This paper is solely motivated by the need for one.

In this paper, we introduce a new formulation for the privacy-utility tradeoff problem and present a custom heuristic algorithm to solve the problem. We capture both privacy and utility using \textit{Mutual Information} and in addition, show that there exists a simple relationship between \textit{Mutual Information} and another metric, \textit{Fisher Information}, which helps understand the fundamental properties of the two metrics and accordingly, their use cases. It should be noted that we do not aim to propose a particular metric to quantify privacy or utility, rather, we focus on capturing the overlooked aspects of the problem. We further discuss the need of a heuristic algorithm motivated by the customizability requirement and present one such algorithm with polynomial time and space complexity.

The remainder of the paper is organized as follows: in Section II, we briefly review closely related works and motivate this paper. In Section III, we present a detailed exposition of the problem setting, formulation and findings. In Section IV, we discuss our heuristic greedy algorithm and evaluate its time and space complexity. In section V, we present experimental results of running our algorithm on simulated datasets. Finally, in Section VI, we close with concluding remarks.

\section{Related Work}
The privacy-utility tradeoff problem has been studied in various settings by several researchers. Li et al. \cite{li2009tradeoff} study the problem in the context of data publishing. They formulate privacy loss as the information gained about the sensitive values of individuals and utility loss as the information lost about the sensitive values of the whole population. The reference for evaluating privacy loss is a completely sanitized data where all quasi-identifiers have been removed and the reference for evaluating utility loss is an unsanitized data. \textit{JS-divergence} and \textit{KL-divergence} measures are used to quantify the privacy loss and the utility loss respectively. Similarly, Makhdoumi et al. \cite{makhdoumi2013privacy} consider the setting where a user wants to release some data in return of some utility. They define privacy loss as the information leaked about some private data from a randomized and disclosed non-private data and quantify it using two metrics: mutual information and maximal leakage. Similarly, they define utility loss as the average distortion between the perturbed and the original data and quantify it using a general distortion measure. We can find a similar model for privacy and utility in the context of smart meters in \cite{rajagopalan2011smart}. 

A more general formulation for privacy can be found in \cite{du2012privacy} where privacy is captured with a generalized cost function with the cost gain measuring the amount of information obtained about the private data after observing the disclosed non-private data. Two privacy metrics,\textit{ average information leakage} and \textit{maximum information leakage} are studied under the self-information cost function. Similarly, utility is quantified as an average distortion in the disclosed data. The privacy-utility trade-off problem is then formulated as a convex optimization problem which involves the minimization of the cost gain subject to utility constraints. The same framework lays the foundation for \cite{makhdoumi2014information} where the log-loss function (self-information) is used for both privacy and utility metrics. Here, the privacy leakage is measured as the mutual information between the private data and the disclosed data and the average distortion (utility) as the mutual information between the non-private data and the disclosed data. The privacy-utility trade-off problem is then formulated as an optimization problem that minimizes the mutual information between the private and the disclosed data over all feasible randomization mechanisms that guarantee the desired distortion level. The problem is referred to as the \textit{Privacy Funnel} and is shown to be non-convex. A greedy algorithm that converges to a solution (potentially a local optimum) is presented to solve the problem.

In \cite{alvim2011differential}, the authors model a query system with an information channel and study the privacy-utility tradeoff. They define privacy as the amount of information leaked about the whole database from the randomized result of queries and quantify it using maximal leakage. Similarly, they define utility as the amount of information about the actual answers obtainable from the randomized answers and quantify it using a gain function that measures the average gain. Likewise, the authors of \cite{sankar2013utility} consider a setting where a database consists of private and public attributes. They define privacy as the entropy of the private attributes conditioned on the knowledge of the randomized and disclosed public attributes and utility as the accuracy of the disclosed data. The privacy leakage is quantified using mutual information and the utility loss via a general distortion function that measures the average distortion of disclosed data. The potential choices for the distortion function include Euclidean distance, Hamming distance or K-L divergence depending on the distribution of the data and characteristics of the database. Similar setting and model for privacy and utility can also be found in \cite{sankar2010theory}.

The existing literature is rich with works that model privacy loss as the information leaked about some private data from the disclosed data and utility loss as the average distortion in the disclosed data due to randomization (for some other works, see \cite{asoodeh2015maximal}, \cite{erdogdu2015privacy}, \cite{wang2017estimation}, \cite{kalantari2017information}, \cite{basciftci2016privacy} and \cite{wang2018utility}). In this paper, we follow a similar model for privacy but devise a different but a practical model of utility and formulate a new privacy-utility tradeoff problem. Our formulation is motivated by two needs: a) to facilitate better privacy-utility tradeoff by associating utility with specific utility attributes b) to account for acceptable loss in utility per unit gain in privacy, or vice-versa. To the best of our knowledge, such a formulation has not yet been discussed in the literature. 

\section{Problem Setup}

\subsection{Problem Setting}
\begin{figure}[!htb]
	\centering
	\includegraphics[width=0.67\columnwidth]{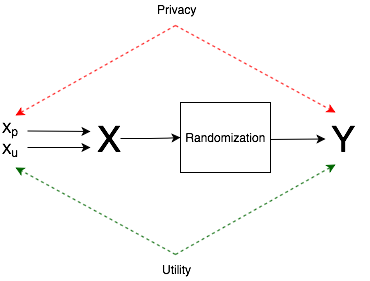}
	\caption{Visual representation of the privacy and the utility aspects}
	\label{fig:priv-util}
\end{figure}

Consider a setting where a user shares some personal information, for instance, on social media, in hope of some utility. In this setting, we first characterize each user by a set of features. Some examples of possible features include gender, political affiliation, like/dislike on a content, rating given to a movie etc. We assume that each user has some \textit{private features} represented by the random vector $X_p$ and some \textit{useful features} represented by the random vector $X_u$. For generality, we do not require that private and useful features be distinct. We, however, assume that a user's utility is computed as a function of the useful features, and hereon, use the term \textit{useful features} and \textit{utility features} interchangeably. Next, we denote by $X$ all the other features that are non-private and non-useful. We, however, assume that $X$ is correlated with $X_p$ and $X_u$ with the goal to release a perturbed version of $X$, say $Y$, that helps gain reasonably large information about $X_u$ but only minimal information about $X_p$. Notice that we are considering a setting where a user does not disclose $X_u$, rather, aims to convey information in $X_u$ by disclosing $Y$. There are at least two reasons for this: first, the exact value of $X_u$ may not be known to the user and second, it may be in the best interest of the user to not disclose $X_u$. The latter case, for instance, is common in social media, where disclosing certain information in hope of gaining gratification may qualify an individual as a narcissist \cite{leung2013generational, sung2016we, liu2016social}. With this setup, we relate privacy inversely to the information gained about $X_p$ from $Y$ and utility directly to the information gained about $X_u$ from $Y$. Note that $(X_p, X_u) \rightarrow X \rightarrow Y$ form a Markov chain. The privacy and the utility aspects of the problem is captured visually in Figure \ref{fig:priv-util}.

\subsection{Problem Formulation}
For our problem formulation, we consider a single user setup. Let $\{x_1, x_2, \cdots x_n\}$ be $n$ random variables representing the actual features of the user. Note that each $x_i$, where $1 \leq i \leq n$, represents a unique feature. In particular, we denote the $n$ features by a single random vector $X$. We also assume that the random variables are correlated with each other. 

In addition to the $n$ features, let the user have two, potentially non-disjoint, sets of special features: a set containing $n_p$ private features and a set containing $n_u$ utility features denoted by the random vectors $X_p$ and $X_u$ respectively. Note that $X_p \cap X = \phi$ and $X_u \cap X = \phi$. However, we assume that $X_p$ and $X_u$ are correlated with $X$. Also, let $X \cup X_p = [x_1, x_2, \cdots, x_n, x^p_{1}, x^p_{2}, \cdots , x^p_{n_p}]^T= [X^T, X_p^T]^T$ and $X \cup X_u = [x_1, x_2, \cdots, x_n, x^u_{1}, x^u_{2}, \cdots , x^u_{n_u}]^T = [X^T, X_u^T]^T$.

Now, let $Y$ represent a perturbed version of $X$. As a randomization mechanism, we assume the addition of Gaussian white noise, i.e. $Y = X + N$, where $N$ represents a random vector with independent components, following a multivariate Gaussian distribution with zero mean and diagonal covariance matrix. Also, let $Y \cup X_p$ and $Y \cup X_u$ denote the perturbed version of $X \cup X_p$ and $X \cup X_u$ respectively. Similarly, let $0^m$ denote a vector of $m$ zeros. Observe that if $\hat N = [N^T, 0^{n_p}]^T$ and $\bar N = [N^T, 0^{n_u}]^T$, then $Y \cup X_p = X \cup X_p + \hat N$ and $Y \cup X_u = X \cup X_u + \bar N$.

For the first part, we quantify both privacy and utility using the mutual information metric. For any random vector $V$, let $\Sigma_{V}$ represent its covariance matrix. The privacy-utility tradeoff problem can now be formulated as the following optimization problem:
\begin{align}
    &minimize\ I(X_p; Y) \nonumber \\
    subject\ to: \nonumber \\
    &I(X_u;X) - I(X_u;Y) \leq \delta, \\
    &\frac{\Delta I(X_p; Y)}{\Delta I(X_u; Y)}  \geq \gamma, \\
    &\Sigma_N \geq 0.
\end{align}

The first constraint enforces a requirement that the end utility loss must be no more than the preset baseline, $\delta$. The second constraint enforces a requirement that the ratio of the end privacy gain to the end utility loss must be no less than the preset threshold, $\gamma$. The third constraint requires that the variance of all added noises be non-negative. Notice that $\delta$ and $\gamma$ are user-customizable parameters. For the sake of both simplicity and clarity, hence forth the first constraint will be referred to as $\delta$-constraint and the second constraint as $\gamma$-constraint. 

For this problem, we assume that $X, X_u, X_p$ are jointly Gaussian. Observe that $\Sigma_Y = \Sigma_X + \Sigma_N$, $\Sigma_{Y \cup X_p} = \Sigma_{X \cup X_p} + \Sigma_{\hat N}$ and $\Sigma_{Y \cup X_u} = \Sigma_{X \cup X_u} + \Sigma_{\bar N}$. Now,
\vspace{2mm}
\begin{align*}
I(X_p; Y) 
    &= H(X_p) + H(Y) - H(X_p,Y) 
    \\
    &= \frac{1}{2}log((2\pi e)^{n_p}|\Sigma_{X_p}|) +
    \frac{1}{2}log((2\pi e)^{n}|\Sigma_{Y}|) \\
    &- \frac{1}{2}log((2\pi e)^{n+n_p}|\Sigma_{Y \cup X_p}|)
    \\
    &= \frac{1}{2}(log|\Sigma_{X_p}| + 
    log|\Sigma_{Y}| - 
    log|\Sigma_{Y \cup X_p}|), \\
\end{align*}
\begin{align*}
I(X_u; Y) 
    &= H(X_u) + H(Y) - H(X_u,Y) 
    \\
    &= \frac{1}{2}(log|\Sigma_{X_u}| + 
    log|\Sigma_{Y}| - 
    log|\Sigma_{Y \cup x_u}|), \\
\\
I(X_u; X) 
    &= H(X_u) + H(X) - H(X_u,X) 
    \\
    &= \frac{1}{2}(log|\Sigma_{X_u}| + 
    log|\Sigma_{X}| - 
    log|\Sigma_{X \cup X_u}|)
\end{align*}
and
\begin{align*}
I(X_u; X) - I(X_u; Y)
    &=
    \frac{1}{2}(log|\Sigma_{X}| - 
    log|\Sigma_{X \cup X_u}| \\
    &-
    log|\Sigma_{Y}| + 
    log|\Sigma_{Y \cup X_u}|). \\
\end{align*}

Observe that the mutual information between two random vectors is a scalar quantity. It measures the total (over all vector components) reduction in uncertainty about a random vector due to the observation of the other random vector.

Before we discuss the intricacies of solving the problem, we will look into another interesting metric that can be used to quantify utility: \textit{Fisher Information}, which measures how much information about a parameter can be obtained by observing a random variable where the probability of the random variable depends on the parameter. The lower bound on the variance of any unbiased estimator of the parameter (formally, referred to as Cramer-Rao bound) is given by the inverse of the Fisher information. Notice that our parameter of interest is the vector of utility attributes and we are interested in measuring the variance of any estimator. In this setting, Fisher information fits perfectly which allows us to formulate the privacy-utility tradeoff problem as follows:

\begin{align*}
    &minimize\ I(X_p; Y) \\
    subject\ to: \\
    &\FISHER(X_u) \geq \delta,   \numberthis \\
    &\frac{\Delta I(X_p; Y)}{\Delta \FISHER(X_u)}  \geq \gamma,  \numberthis \\
    &\Sigma_N \geq 0. \numberthis
\end{align*}

We make the same assumptions about the distribution of $X$, $X_p$ and $X_u$ we made earlier and follow the same notations for all relevant covariance matrices. In addition, we denote by $M_u$ and $M_{Y \cup X_u}$ the mean vectors of $X_u$ and $Y \cup X_u$, respectively. Note that all differentiations, henceforth, are with respect to $X_u$ unless otherwise stated.

\vspace{2mm}
Now,

\begin{align*}
\FISHER(X_u)
&= -\EXPECT[\pdv[2]{\ell(Y|X_u)}{X_u}] \\
&= -\EXPECT[\ell''(Y|X_u)].   \numberthis \label{eqn:I(Xu)}
\end{align*}

Here,
\begin{align*}
\ell(Y|X_u) &= 
\log f(Y|X_u) \\
&= \log f(Y, X_u) - \log f(X_u).
\end{align*}

Differentiating both sides, we get
\begin{align*}
\ell'(Y|X_u) &= 
\frac{f'(Y,X_u)}{f(Y,X_u)} - \frac{f'(X_u)}{f(X_u)}. \numberthis \label{eqn:ell'}
\end{align*}

We have
\begin{align*}
f(X_u) &=
    \!\begin{aligned}[t]
    &\frac{1} {(2\pi)^{n_u/2}|\Sigma_{X_u}|^{1/2}} \\
    &\cdot \exp(-\frac{1} {2}
    (X_u - M_u)^T
    \Sigma_{X_u}^{-1} (X_u - M_u)). 
    \end{aligned}
\\
\end{align*}

Differentiating both sides,
\begin{align*}
f'(X_u) &=
        \!\begin{aligned}[t]
        &\frac{1} {(2\pi)^{n_u/2}|\Sigma_{X_u}|^{1/2}} \\
        &\cdot \exp(-\frac{1} {2}
        (X_u - M_u)^T
        \Sigma_{X_u}^{-1} (X_u - M_u)) \\
        &\cdot
        \frac{\partial}{\partial X_u} \Bigr(-\frac{1} {2} (X_u - M_u)^T \Sigma_{X_u}^{-1} (X_u - M_u) \Bigl)
        \end{aligned}
\\
    &=
    f(X_u) \cdot
    \frac{\partial}{\partial X_u} \Bigr(-\frac{1} {2} (X_u - M_u)^T \Sigma_{X_u}^{-1} (X_u - M_u) \Bigl). 
\\
\end{align*}

Similarly,
\begin{align*}
f(Y, X_u) &=
    \!\begin{aligned}[t]
    &\frac{1} {(2\pi)^{(n+n_u)/2}|\Sigma_{Y \cup X_u}|^{1/2}} \\
    &\cdot \mathrm{exp}\Bigl( 
        \!\begin{aligned}[t]
        &-\frac{1} {2} (Y \cup X_u - M_{Y \cup X_u})^T \\
        &\cdot \Sigma_{Y \cup X_u}^{-1} (Y \cup X_u - M_{Y \cup X_u}) \Bigr). 
        \end{aligned}
    \end{aligned}
\end{align*}

Differentiating both sides,
\begin{align*}
f'(Y, X_u) &=
    \!\begin{aligned}[t]
    &f(Y, X_u) \\
    &\cdot \frac{\partial}{\partial X_u} \Bigl(
        \!\begin{aligned}[t]
        &-\frac{1} {2} (Y \cup X_u - M_{Y \cup X_u})^T \\
        &\cdot \Sigma_{Y \cup X_u}^{-1} (Y \cup X_u - M_{Y \cup X_u}) \Bigr). 
        \end{aligned}
    \end{aligned}
\end{align*}

Now, (\ref{eqn:ell'}) can be written as
\begin{align*}
\ell'(Y|X_u) &= 
    \frac{\partial}{\partial X_u} \Bigl(
        \!\begin{aligned}[t]
        &-\frac{1} {2} (Y \cup X_u - M_{Y \cup X_u})^T \\
        &\cdot \Sigma_{Y \cup X_u}^{-1} (Y \cup X_u - M_{Y \cup X_u})\Bigr) 
        \end{aligned} 
        \\
    &- \frac{\partial}{\partial X_u} \Bigl(-\frac{1} {2} (X_u - M_u)^T \Sigma_{X_u}^{-1} (X_u - M_u) \Bigr). 
\end{align*}

Differentiating both sides,
\begin{align*}
\ell''(Y|X_u) &= 
    \frac{\partial^2}{\partial X_u^2} \Bigl(
        \!\begin{aligned}[t]
        &-\frac{1} {2} (Y \cup X_u - M_{Y \cup X_u})^T \\
        &\cdot \Sigma_{Y \cup X_u}^{-1} (Y \cup X_u - M_{Y \cup X_u})\Bigr) 
        \end{aligned}
        \\
    &- \frac{\partial^2}{\partial X_u^2} \Bigl(-\frac{1} {2} (X_u - M_u)^T \Sigma_{X_u}^{-1} (X_u - M_u) \Bigr). \numberthis
    \label{eqn:ell''}
\end{align*}

Let $a_{i,j}$ and $b_{i,j}$ denote the elements in the $i^{th}$ row and $j^{th}$ column of the inverse matrices, $\Sigma_{X_u}^{-1}$ and $\Sigma_{Y \cup X_u}^{-1}$, respectively. Then, (\ref{eqn:ell''}) simplifies to a Hessian matrix, $H$, with dimension $n_u \cross n_u$ where

\begin{align*}
H_{i,j} &=
    \frac{1}{2} \cdot
    \begin{cases} 
        2(a_{i,i} - b_{n+i,n+i}) & i = j \\
        a_{i,j}+a_{j,i}-b_{n+i,n+j}-b_{n+j,n+i} & i \ne j
   \end{cases}
\end{align*}

Therefore, (\ref{eqn:I(Xu)}) can be written as
\begin{align*}
  \FISHER(X_u) &= -\EXPECT(H) = H,   \numberthis \label{eqn:I(Xu):2}
\end{align*}

where the negative expectation of the Hessian is commonly referred to as the \textit{Fisher Information Matrix}.

The expression in (\ref{eqn:I(Xu):2}) reveals interesting facets of the utility formulation using Fisher information. Recall that the inverse of the Fisher information gives the lower bound on the variance of any unbiased estimator of a parameter. The inverse of the Fisher information matrix therefore, gives the estimates of the variances and covariances of $n_u$ estimators of the $n_u$ utility parameters. An important takeaway is that the Fisher information formulation of utility accounts for the utility associated with each utility attribute unlike the mutual information formulation which averages out the utility across all utility attributes. The choice of one metric over the other for capturing utility therefore, depends on whether individual utility needs to be accounted for or whether capturing the average utility is sufficient. 

A special case of (\ref{eqn:I(Xu):2}) is when there is a single utility feature ($n_u = 1$). In this case, the Hessian matrix is a scalar with the first and only element given by

\begin{align*}
H_{1,1} &= 
    \frac{1}{2} \cdot 2(\frac{1}{\sigma_{x_u}^2} - b_{n,n})
    \\
    &=
    \frac{1}{\sigma_{x_u}^2} - b_{n,n}.
\end{align*}

Then, from (\ref{eqn:I(Xu):2}),
\begin{align*}
\FISHER(X_u) &=
    b_{n,n} - \frac{1}{\sigma_{x_u}^2},  \numberthis \label{eqn:I(X_u):3}
\end{align*}
where $\sigma_{x_u}^2$ represents the variance of the utility feature.

\subsection{Relationship between Mutual Information and Fisher Information}
Expanding on the expression given in (\ref{eqn:I(X_u):3}), we now establish a relationship between Mutual information and Fisher information for $n_u = 1$. First, observe that

\begin{align*}
b_{n,n} &=
    \frac{|\Sigma_Y|} {|\Sigma_{Y \cup X_u}|}.
\end{align*}

Multiplying both sides by $\sigma_{x_u}^2$ and computing $\frac{1}{2}\log$ on both sides, we get
\begin{align*}
\frac{1}{2} \log(b_{n,n} \cdot \sigma_{x_u}^2) &=
    \frac{1}{2} \log( \frac{\sigma_{x_u}^2 \cdot |\Sigma_Y|} {|\Sigma_{Y \cup X_u}|} ) 
    \\
    &= I(X_u;Y). \numberthis \label{eqn:I(Xu;Y)}
\end{align*}

Then, from (\ref{eqn:I(X_u):3}) and (\ref{eqn:I(Xu;Y)}),
\begin{align*}
    I(X_u;Y) &= 
    \frac{1}{2} log(\sigma_{x_u}^2 \cdot \FISHER(X_u) + 1).  \numberthis \label{eqn:I(X_U;Y):2}
\end{align*}

Observe that the expression in (\ref{eqn:I(X_U;Y):2}) is analogous to the expression for \textit{channel capacity} in communication systems, where $I(X_u;Y)$ corresponds to the channel capacity,  $\sigma_{x_u}^2$ corresponds to the signal power and $\frac{1}{\FISHER(X_u)}$ corresponds to the noise power.

\section{A Heuristic approach to solving the privacy-utility tradeoff problem}
The optimization problem formulated in section III.B 
cannot be readily solved using existing methods due to the additional $\gamma-$constraint. Furthermore, the convexity of the objective function is not known which makes the existing convex optimization techniques inapplicable. Analytical methods based on KKT conditions that restrictively work on certain non-convex problems also fall short for two reasons: first, it is unknown whether strong duality holds for the problem and second, these methods often do not scale well for higher dimensional problems. These restrictions motivate us to develop a custom heuristic algorithm to solve the problem.

We take a greedy iterative approach to solving the problem: at each step, we add a small amount of noise, say $\Delta\theta$, to one of the variables, $x_1, x_2, \cdots, x_n$. The selection of the variable to add noise to is determined by the \textit{gain factor} which is defined as the ratio of the \textit{privacy gain} to the \textit{utility loss} due to the added noise. Essentially, in each iteration, we select the variable with the highest gain factor, add $\Delta\theta$ amount of noise to the variable and test for the $\delta$ and the $\gamma$ constraints. If both constraints are slack, we commit with the noise-addition, else, we reduce $\Delta\theta$ by a factor of 2 and proceed to the next iteration without committing. We stop when $\Delta\theta$ is less than or equal to a preset value $\epsilon$.

An interesting situation arises when a variable yields the highest gain factor among all variables but only achieves negligible privacy gain (as a result of a small utility loss). Clearly, it is not worthwhile to add any more noise to the variable. This is called a \textit{saturation} phase and the variable is said to be \textit{saturated}. If a variable is saturated, we ignore the variable for the current iteration and continue with the other variables. If all variables are saturated in the same iteration, this is referred to as \textit{total saturation}, in which case, we stop.

In what follows, we summarize our approach by presenting a heuristic greedy algorithm. For simplicity, let $Y_i + \Delta\theta$ represent the vector that has the same elements as $Y$ but with $\Delta\theta$ amount of noise added to the $i^{th}$ component. 

\vspace{2mm}
\textbf{Greedy algorithm for the privacy-utility tradeoff problem:}

\begin{enumerate}
    \item \textbf{Initialization}. Initialize $\Delta\theta$ to a small positive value, set Y = X.
    \item \textbf{Evaluation}. For each variable, $i$ $(1 \leq i \leq n)$, compute \\
    \textbullet\ privacy\_gain($i$) $= I(X_p; Y) - I(X_p; Y_i+\Delta\Theta)$ \\
    \textbullet\ utility\_loss($i$) $= I(X_u; Y) - I(X_u; Y_i+\Delta\Theta)$ \\
    \textbullet\ gain\_factor($i$) $=$ privacy\_gain($i$) / utility\_loss($i$) \\
    \textbullet\ If privacy\_gain($i$) $< \epsilon_0$, set gain\_factor($i$) $= -1$ \\
    (This corresponds to the saturation phase)
    \item \textbf{Selection}. Select the variable with the highest gain\_factor. Let $j$ be the index of this variable. 
    \item \textbf{Stopping criteria}. If gain\_factor($j$) $<= 0$, stop. \\
    (If the highest gain factor $<=0$, all other gain factors are also $<=0$ which implies total saturation)
    \item \textbf{Update}. If  $I(X_u; Y_j+\Delta\Theta) \leq \gamma$, set $Y = Y_j + \Delta\theta$. Else, set $\Delta\theta = \Delta\theta/2$.
    \item \textbf{Repeat}. If $\Delta\theta < \epsilon$, stop. Else, go to 2.
\end{enumerate}

\vspace{2mm}
Although in the above algorithm, we have quantified both privacy gain and utility loss using Mutual information, it is straightforward to modify the algorithm and quantify utility loss using Fisher information. However, note that the values of $\delta$ and $\gamma$ need to be adjusted for the new metric. The new values of $\delta$ and $\gamma$ can easily be determined using the relationship in (\ref{eqn:I(X_U;Y):2}).

\subsection{Algorithmic Complexity}
There are two relatively computationally intensive parts of the algorithm:
    \begin{enumerate}
        \item Computing the Mutual information
        \item Determining the variable with the highest gain factor
    \end{enumerate}
    
Computing the Mutual information requires computing the determinants of the covariance matrices. The dimensions of the largest matrix is $(n+1) \cross (n+1)$ which requires, roughly, $O(n^3)$ time to compute its determinant. Similarly, determining the variable with the highest gain factor requires sorting which takes, on average, $O(n\log(n))$ time. Overall, the time-complexity of the algorithm is $O(n^3)$.
    
In regard to the space-complexity, there are two relatively space intensive parts of the algorithm:
    \begin{enumerate}
        \item Storing the covariance matrices
        \item Storing the gain factors of $n$ variables
    \end{enumerate}

The space requirement for the covariance matrices is in the order of $O(n^2)$. Similarly, the space requirements for storing the gain factors of $n$ variables is in $O(n)$. The overall space-complexity of the algorithm is, therefore, $O(n^2)$.

\section{Experimental Analysis}
In this section, we analyze our model by running the greedy algorithm on simulated datasets. All datasets are sampled from a multivariate normal distribution and reflect various features of a user. The covariance matrices for the sample datasets and the corresponding privacy-utility trade-off graphs are presented below:

\vspace{2mm}
\noindent\textbf{Sample dataset 1:}
\begin{align*}
    \Sigma_X &= 
    \begin{bmatrix}
    138.27 & 165.66 \\
    165.66 & 240.07 \\
    \end{bmatrix},\  
    \\
    \Sigma_{X \cup X_p} &= 
    \begin{bmatrix}
    138.27 & 165.66 & 26.36 \\
    165.66 & 240.07 & 43.86 \\
    26.36 & 43.86 & 8.76 \\
    \end{bmatrix},\ 
    \\
    \Sigma_{X \cup X_u} &= 
    \begin{bmatrix}
    138.27 & 165.66 & 11.28 \\
    165.66 & 240.07 & 6.84 \\
    11.28 & 6.84 & 2.26 \\
    \end{bmatrix}
\end{align*}

\noindent\textbf{Sample dataset 2:}
\begin{align*}
    &\Sigma_X = \\ 
    &\begin{bmatrix}
    66.42 & 57.38 & 83.90 & 80.03 & 0.06 & 121.43 \\
    57.38 & 229.20 & 146.94 & 232.62 & 0.04 & 69.30 \\
    83.90 & 146.94 & 142.89 & 169.83 & 0.06 & 140.22 \\
    80.03 & 232.62 & 169.83 & 247.38 & 0.06 & 114.44 \\
    0.06 & 0.04 & 0.06 & 0.06 & 0.12 & 0.10 \\
    121.43 & 69.30 & 140.22 & 114.44 & 0.10 & 233.30 \\
    \end{bmatrix},\  
    \\ \\
    &\Sigma_{X \cup X_p} = \\ 
    &\begin{bmatrix}
    66.42 & 57.38 & 83.90 & 80.03 & 0.06 & 121.43 & 9.26 \\
    57.38 & 229.20 & 146.94 & 232.62 & 0.04 & 69.30 & 45.07 \\
    83.90 & 146.94 & 142.89 & 169.83 & 0.06 & 140.22 & 27.17 \\
    80.03 & 232.62 & 169.83 & 247.38 & 0.06 & 114.44 & 45.18 \\
    0.06 & 0.04 & 0.06 & 0.06 & 0.12 & 0.10 & 0.01 \\
    121.43 & 69.30 & 140.22 & 114.44 & 0.10 & 233.30 & 9.44 \\
    9.26 & 45.07 & 27.17 & 45.18 & 0.01 & 9.44 & 9.01 \\
    \end{bmatrix}\ 
    \\ \\
    &\Sigma_{X \cup X_u} = \\ 
    &\begin{bmatrix}
    66.42 & 57.38 & 83.90 & 80.03 & 0.06 & 121.43 & 11.22 \\
    57.38 & 229.20 & 146.94 & 232.62 & 0.04 & 69.30 & 2.42 \\
    83.90 & 146.94 & 142.89 & 169.83 & 0.06 & 140.22 & 11.31 \\
    80.03 & 232.62 & 169.83 & 247.38 & 0.06 & 114.44 & 6.93 \\
    0.06 & 0.04 & 0.06 & 0.06 & 0.12 & 0.10 & 0.01 \\
    121.43 & 69.30 & 140.22 & 114.44 & 0.10 & 233.30 & 22.38 \\
    11.22 & 2.42 & 11.31 & 6.93 & 0.01 & 22.38 & 2.23 \\
    \end{bmatrix}
\end{align*}

\begin{figure}[!htb]
	\centering
	\includegraphics[width=1.0\columnwidth]{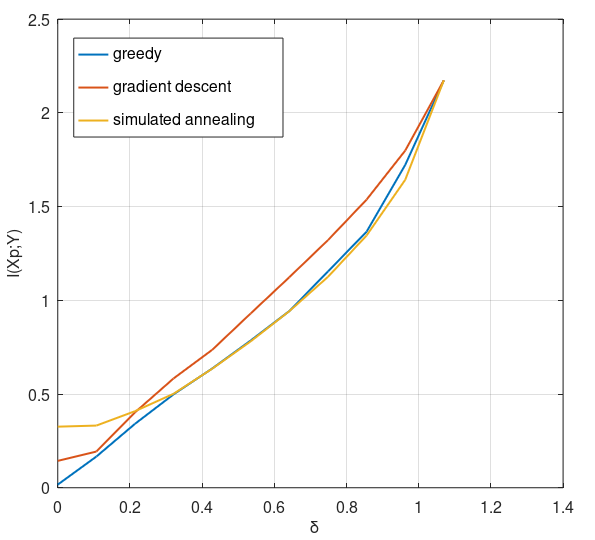}
	\caption{Performance of different algorithms on sample dataset 1; the graph shows the plot of minimum $I(X_p;Y)$ for different values of $\delta$ with $\gamma$ = 0 ($\epsilon_0 =  10^{-6}$)}
	\label{fig:plot1a}
\end{figure}

\begin{figure}[!htb]
	\centering
	\includegraphics[width=1.0\columnwidth]{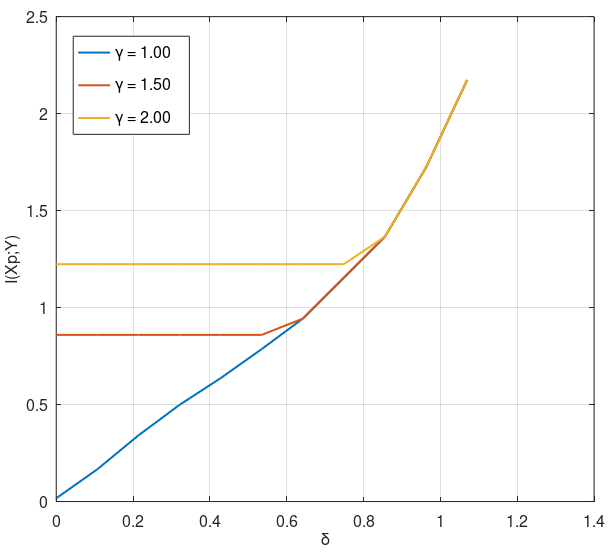}
	\caption{Simulation of the greedy algorithm on sample dataset 1; the graph shows the plot of minimum $I(X_p;Y)$ for different values of $\delta$ and $\gamma$ ($\epsilon_0 = 10^{-6}$)}
	\label{fig:plot1b}
\end{figure}

\begin{figure}[!htb]
	\centering
	\includegraphics[width=1.0\columnwidth]{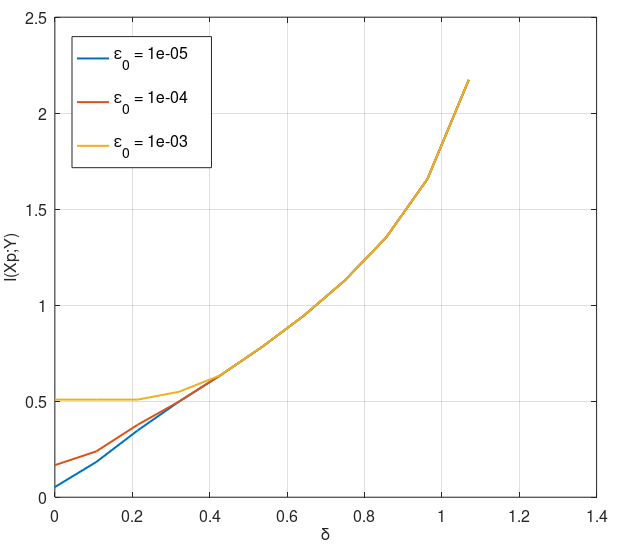}
	\caption{Performance of the greedy algorithm on sample dataset 1 for different values of $\delta$ and $\epsilon_0$ ($\gamma = 0$)}
	\label{fig:plot1c}
\end{figure}

\begin{figure}[!htb]
	\centering
	\includegraphics[width=1.0\columnwidth]{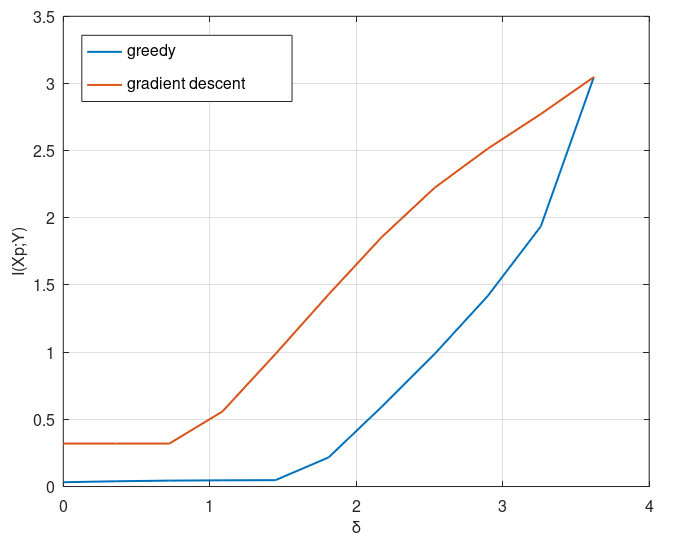}
	\caption{Performance of the greedy and gradient descent algorithms on sample dataset 2; the graph shows the plot of minimum $I(X_p;Y)$ for different values of $\delta$ with $\gamma$ = 0 ($\epsilon_0 =  10^{-6}$)}
	\label{fig:plot2a}
	\vspace{1mm}
\end{figure}

\begin{figure}[!htb]
	\centering
	\includegraphics[width=1.0\columnwidth]{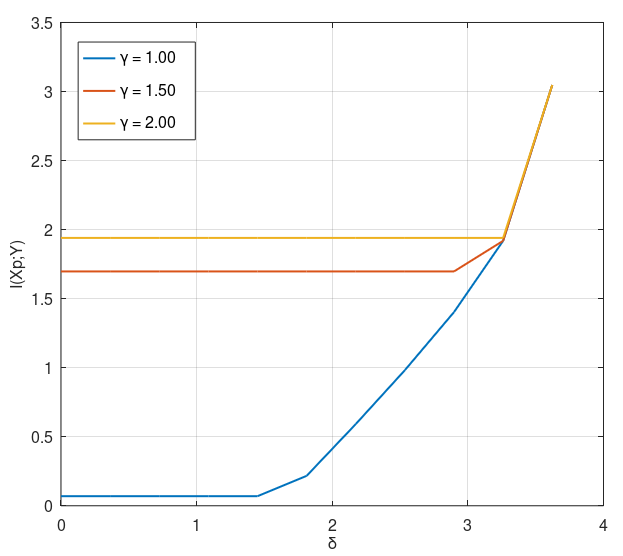}
	\caption{Simulation of the greedy algorithm on sample dataset 2; the graph shows the plot of minimum $I(X_p;Y)$ for different values of $\delta$ and $\gamma$ ($\epsilon_0 =  10^{-6}$)}
	\label{fig:plot2b}
\end{figure}

\begin{figure}[!htb]
	\centering
	\includegraphics[width=1.0\columnwidth]{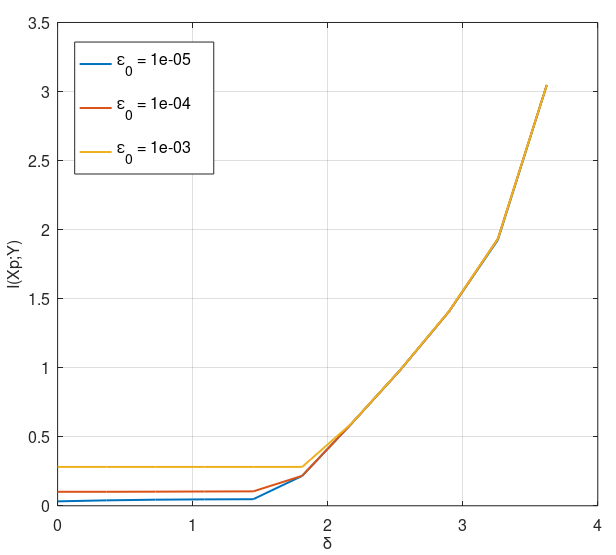}
	\caption{Performance of the greedy algorithm on sample dataset 2 for different values of $\delta$ and $\epsilon_0$ ($\gamma = 0$)}
	\label{fig:plot2c}
\end{figure}

Figure \ref{fig:plot1a} compares the performance (in terms of minimizing the objective function) of the greedy algorithm against the gradient descent and simulated annealing algorithms for sample dataset 1.\footnote{A simple gradient descent algorithm that numerically approximates the gradient was used for the simulation. Simulations of the gradient descent and the simulated annealing algorithms involved the addition of quadratic loss functions to the objective function to transform the constrained optimization problem into an equivalent unconstrained optimization problem.} Similarly, Figure \ref{fig:plot2a} compares the performance of the greedy algorithm against the gradient descent algorithm for sample dataset 2.\footnote{The complexity of designing a robust neighbor function for the higher dimensional problem hindered us from running the simulated annealing algorithm on sample dataset 2.} As the gradient descent algorithm is not compatible with the $\gamma$-constraint, for the sake of comparison, we set $\gamma = 0$ for our simulations (which is equivalent to omitting the $\gamma$-constraint). Observe in the two figures that our greedy algorithm consistently performs better than the gradient descent algorithm across different values of delta. Where applicable, the results are comparable to that of the simulated annealing algorithm. We note that the figures are not meant to highlight the superiority of our algorithm\footnote{Gradient descent algorithms are typically faster and can be optimized for better accuracy.} but to show that our algorithm converges to a reasonably good solution. We, however, stress that the added $\gamma-$constraint in the original problem often mandates the use of our algorithm (or a modified version of it).

Figure \ref{fig:plot1b} and Figure \ref{fig:plot2b} show the minimum values of the objective function, $I(X_p;Y)$, across different values of $\delta$ and $\gamma$. As can be seen in the figures, both $\delta$ and $\gamma$ parameters determine the end mutual information between $X_p$ and $Y$, and the corresponding privacy gain.  For $\gamma = 0$, the privacy gain is higher for lower values of $\delta$ (to the point of saturation) as a result of lower mutual information between $X_p$ and $Y$ as compared to the higher values of $\delta$. However, when $\gamma \neq 0$, the end privacy gain may be the same over a range of $\delta$ values. Note that these inferences are consistent with our intuitive understanding of the functionalities of the $\delta$ and $\gamma$ parameters: $\gamma = 0$ implies that the user does not care about the gain in privacy per unit loss in utility and therefore, we expect the end privacy gain to be solely dependent on the desired level of utility loss, $\delta$. However, $\gamma \neq 0$ implies that the gain in privacy per unit loss in utility must also be prioritized when maximizing the privacy gain (by minimizing the objective function). Under such constraints, we expect the smaller values of $\delta$ to be less relevant in determining the minimum $I(X_p;Y)$ as the user demands higher gain factors.

To further understand the effect of the $\delta$ and the $\gamma$ parameters on the minimum value of the objective function, the graphs shown in Figure \ref{fig:plot1b} and Figure \ref{fig:plot2b} can be virtually divided into two regions, a $\delta$-dominated region and a $\gamma$-dominated region. For a given $\gamma$, if the minimum $I(X_p;Y)$ is constant across a range of $\delta$ values (not accounting for the saturation), we say that the corresponding region is $\gamma$-dominated. Similarly, for a given $\gamma$, if the minimum $I(X_p;Y)$ varies with $\delta$, we say that the corresponding region is $\delta$-dominated. Note that while it may appear that larger values of $\gamma$ results in larger $\gamma$-dominated region and smaller values of $\gamma$ results in smaller $\gamma$-dominated region, the extent of the region significantly depends on the characteristics of the user attributes as well. Consider, for instance, two attributes of a user, $X_1$ and $X_2$, where both $X_1$ and $X_2$ are highly correlated with the private attribute, $X_p$, and less correlated with the utility attribute, $X_u$. The optimal privacy mechanism intuitively involves adding a lot of noise to both $X_1$ and $X_2$ without losing much utility. Essentially, the gain factor is expected to be very high throughout the addition of incremental noises. For this setup, higher values of $\gamma$ (up to a threshold) do not necessarily imply larger $\gamma$-dominated regions. Put simply, the extent of the $\gamma$-dominated and the $\delta$-dominated regions depends on the covariances of the attributes as much as the parameters themselves.

Figure \ref{fig:plot1c} and Figure \ref{fig:plot2c} highlight the sensitivity of our algorithm to the $\epsilon_0$ parameter. The $\epsilon_0$ parameter defines the threshold for saturation and consequently, influences the resulting solution. For smaller values of $\delta$, observe that the objective function is more sensitive to the $\epsilon_0$ parameter. Also observe that the smaller values of $\epsilon_0$ consistently produce smaller minimum values for $I(X_p;Y)$ across all values of $\delta$ and are therefore, desirable. However, we note that for smaller values of $\epsilon_0$, the algorithm converges more slowly.

Running the greedy algorithm on the two datasets, in the case in which the utility is expressed as the Fisher information and $\delta$ adjusted accordingly as in (\ref{eqn:I(X_U;Y):2}), yields exactly the same privacy vs utility curves as above. This identity was consistently observed through all our simulations. Nevertheless, because the problem is potentially non-convex, we conjecture that under certain instantiations the two utility metrics will provide different minimum privacy values.

\section{Conclusion}
One of the fundamental challenges in sharing data online is ensuring that the released data does not leak private information while, at the same time, yields maximum utility. Seeking absolute privacy and maximum utility of the shared data are contradictory goals as there is always a trade-off involved. Commonly known as the privacy-utility trade-off problem, we find different formulations of the problem in the literature with different application-specific solutions. However, there are some aspects of the problem that had not yet been captured in the existing works. In this paper, we captured those aspects by introducing a new formulation of the problem. In particular, we modelled utility differently and discussed the possibility of using different metrics to quantify utility under our model. We established a simple relationship between two candidate metrics, \textit{Mutual Information} and \textit{Fisher Information}, albeit in a restricted setting. We further introduced a custom greedy algorithm with polynomial time and space complexity to solve the problem and presented experimental results.
{\Large \pitslab}

\balance
\bibliographystyle{plain}
\bibliography{references}

\end{document}